\documentclass{emulateapj}

\defcitealias{zag10}{ZBW10} 	
\defcitealias{fer06}{FBW06} 	
\defcitealias{rob06}{ROB06} 	

\shorttitle{Gas braking in the $\beta$~Pic disk}
\shortauthors{Brandeker}

\begin{document}

\title{Exposing the gas braking mechanism of the
$\beta$~Pictoris disk$^{*}$}\thanks{$^{*}$ Based on observations made
with ESO Telescopes at the La Silla Observatory.}

%\title{Exposing the gas braking mechanism of the
%$\beta$~Pictoris disk\footnote{Based on observations made
%with ESO Telescopes at the La Silla Observatory}}

\author{Alexis Brandeker}
\affil{Dept.\ of Astronomy, Stockholm University, SE-106\,91 Stockholm, Sweden}

\begin{abstract}
Ever since the discovery of the edge-on circumstellar disk around
$\beta$~Pictoris, a standing question has been why the gas observed
against the star in absorption is not rapidly expelled by the strong
radiation pressure from the star. A solution to the puzzle has been
suggested to be that the neutral elements that experience the radiation
force also are rapidly ionized, and so are only able to accelerate
to an average limiting velocity $v_{\mathrm{ion}}$. Once ionized,
the elements are rapidly braked by \ion{C}{2}, which is observed to be
at least 20$\times$ overabundant in the disk with respect to other
species. A prediction from this scenario is that different neutral
elements should reach different $v_{\mathrm{ion}}$, depending on
the ionization thresholds and strengths of driving line transitions.
In particular, neutral Fe and Na are predicted to reach the radial velocities
0.5 and 3.3\,km\,s$^{-1}$, respectively, before being ionized. In this paper
we study
the absorption profiles of Fe and Na from the circumstellar gas disk
around $\beta$\,Pic, as obtained by HARPS at the ESO 3.6\,m telescope.
We find that the Fe and Na velocity profiles are indeed shifted
with respect to each other, confirming the model. The absence of an
extended blue wing in the profile of Na, however, indicates that there
must be some additional braking on the neutrals. We explore the possibility
that the ion gas (dominated by \ion{C}{2}) can brake the neutrals, and
conclude that about 2--5$\times$ more C than previously estimated is
needed for the predicted line profile to be consistent with the observed
one.
\end{abstract}


\section{Introduction}

Understanding the evolution of gas in circumstellar (CS) disks is critical
to understand star and planet formation. Unfortunately, CS
gas in general is very difficult to observe, so how and when the gas
disappears, is still poorly known. The main difficulty is that the
cool gas in disks only thermally excites the lowest lying energy levels,
which emit radiation in mid- to far-infrared wavelengths that are
blocked by the terrestrial atmosphere. Radiatively excited levels
are easily observed from the ground, but require very high contrast
sensitivity, as the star typically emits many orders of magnitude
more radiation at the same wavelengths. In the case the CS
disk is seen edge on, however, the gas can be seen in absorption against
the star. This is the case for the disk around $\beta$~Pictoris,
where CS absorption lines were found \citep{sle75} long before
the disk was discovered \citep{aum85,smi84}. $\beta$\,Pic
is a nearby \citep[19.44$\pm$0.05\,pc;][]{van07} A5 main-sequence star
estimated to be 10--20\,Myr old \citep{zuc01,men08}. Because time scales for dust
removal mechanisms are much shorter than the age of the system \citep{bac93},
the dust disk must be replenished. The most popular mechanism to
replenish the dust disk is through collisional cascades from planetesimal
size objects, hence the classification of the $\beta$\,Pic disk
as a \emph{debris} disk.

Even though the fate of CS gas is much less constrained
than the dust, also the gas disk around $\beta$\,Pic is thought
to be secondary as opposed to a remnant from the star forming nebula,
as indicated by the presence of CO \citep{vid94} that quickly is
dissociated \citep{dis88}, and by dynamical arguments \citep*[hereafter \citetalias{fer06}]{fer06}.
Suggested mechanisms for replenishing the gas are photo-desorption
of dust \citep{che07}, evaporation of comets \citep{beu89}, and collisional evaporation
of dust grains \citep{cze07}.

After the discovery of the $\beta$\,Pic CS dust disk,
more attention was directed towards the gas observed in absorption
against the star \citep{hob85}, and it was soon realized that the expelling
radiation force on many elements exceed the gravity --- by up to a factor
of several hundred in the case of \ion{Na}{1}. Yet, most of the gas absorption
is observed to be at rest with respect to the star. Evidently, there
must be some braking agent at work, but the lack of spatial information
of the location of the gas made it difficult to constrain braking
models --- e.g.\ \citet{lag98} suggested that a dense ring of \ion{H}{1}
at sub-AU distance from the star would be able to brake the gas through
friction, without violating any observational constraints known at
the time. This changed with the detection of spatially resolved \ion{Na}{1}
emission from the disk, observed to be co-located with the dust disk
out to hundreds of AU \citep{olo01}. For H to brake the \ion{Na}{1} over the
whole disk, \citet{bra04} estimated that the required mass would
have to be higher than upper limits set on \ion{H}{1} \citep{fre95} and H$_{2}$
\citep{lec01}. 

The problem of how the gas around $\beta$\,Pic is braked was investigated
in detail by \citetalias{fer06}. One key observation they made was that
all neutrals that are subject to a strong radiation force, are also
quickly ionized. The neutrals thus only reach small velocities $v_{\mathrm{ion}}$
on average (depending on element) before being ionized. Since ions
are affected by long-range Coulomb forces, they interact strongly
and act as a single fluid; ions experiencing radiation force are efficiently
braked by those unaffected by radiation. For ions, \ion{C}{2} turns out
to be the most dominant braking agent, while \ion{Fe}{2} is the strongest
driver. For gas of cosmic abundance, the effective radiation force
acting on the gas as a whole, was found to be 4$\times$ stronger
than gravity. In effect, \ion{Fe}{2} would drag the rest of the gas along
and expel it from the system. In order for the ion gas to be stable against
the radiation pressure, an overabundance of at least 10$\times$ of
C would be required, and in fact, FUSE observations of C in absorption
against the star subsequently proved C to be at least 20$\times$
overabundant, providing the stabilizing inertia to brake the disk
\citep[hereafter \citetalias{rob06}]{rob06}.

A prediction from the scenario that only ions are braked, is that
neutrals should be observed to reach a radial velocity of $v_{\mathrm{ion}}$
with respect to the star. The purpose of this paper is to test that
prediction in detail, by making use of archival high spectral resolution
data that has been obtained of $\beta$\,Pic for other purposes.
Since the absolute heliocentric velocity of $\beta$\,Pic is very
hard to measure to good accuracy we treat it as a free parameter and
focus on the observed differential velocity between \ion{Fe}{1} and \ion{Na}{1}.
The paper is organized as follows. In \S\,\ref{sec:theory} we study
what the implications of the braking scenario look like in detail,
in particular the expected absorption line profiles. \S\,\ref{sec:Data}
reports on how the data we use to compare with theory were acquired
and handled, especially the somewhat elaborate procedure to remove
telluric lines from the \ion{Na}{1} region. In \S\,\ref{sec:results} we show
the resulting comparisons, and discuss implications. The paper is
concluded by \S\,\ref{sec:Conclusions}.

\section{Line profiles}
\label{sec:theory}

If disk braking operates on ions, but not on neutrals, we can expect
the neutrals to be accelerated for a short while before getting ionized.
Since the number of accelerating photons outnumber the ionizing photons
by at least two orders of magnitude, we can treat acceleration as
a continuous process while the time before ionization is stochastic
according to an exponential distribution. Given an ensemble of particles,
this will give rise to a distribution of velocities at any given moment,
which will broaden an absorption line and result in a characteristic
profile. In the following two subsections we first study the case
that there is no braking acting on the neutral particle, and then
continue with the case where some braking is provided by a surrounding
ion gas.

\subsection{Freely accelerated neutrals}
\label{sub:free}

To find out what kind of absorption profile we should expect from neutral
elements being radially accelerated, we make the following assumptions: 
\begin{enumerate}
\item The neutral particles are ionized at a rate $\Gamma(r)\propto r^{-2}$,
where $r$ is the distance to the star. The particle has no {}``memory'';
the lifetime $\Lambda$ of a neutral particle is thus exponentially
distributed with the expected lifetime $\Gamma^{-1}$. 
\item The particles are in ionization equilibrium, meaning that the recombination
rate equals the ionization rate. 
\item The particles travel a very short distance before being ionized; $r$ is 
thus essentially constant during the neutral phase.
\item Ionized particles are braked efficiently and so are at zero radial
velocity. Neutral particles thus start with no radial velocity. 
\end{enumerate}
The probability density distribution of the lifetime $\Lambda$ before
ionization is thus 
\begin{equation}
f_{\Lambda}(t)=\Gamma\exp(-\Gamma t).\label{eq:lifetime}
\end{equation}
The radial equation of motion for a particle when it becomes neutral is
\begin{equation}
m\frac{\mathrm{d}v}{\mathrm{d}t}=(\beta-1)\frac{GM_{\star}m}{r^{2}}+\frac{mL^2}{r^3},
\end{equation}
where $\beta=F_{\mathrm{rad}}/F_{\mathrm{grav}}$ is the ratio
between radiation force $F_{\mathrm{rad}}$ and gravity $F_{\mathrm{grav}}$,
$G$ is the gravitational constant, and $M_{\star}$ is the mass of the central star.
The angular momentum $L = (GMr_0)^{1/2}$, assuming the particle becomes neutral
while in a circular Keplerian orbit of radius $r_0$. For a particle moving a
short radial distance before becoming ionized again (and thus $r\approx r_0$), 
the radial acceleration during its neutral lifetime can be approximated with
\begin{equation}
a=\beta\frac{GM_{\star}}{r^{2}},
\end{equation}
The velocity reached by a particle before ionization
is then $V_{\mathrm{ion}}=a\Lambda$, and its velocity probability density
distribution \begin{equation}
f_{V_{\mathrm{ion}}}(v)=v_{\mathrm{ion}}^{-1}\exp(-\frac{v}{v_{\mathrm{ion}}}),\label{eq:profile}\end{equation}
 where we have substituted $v\equiv at$ and $v_{\mathrm{ion}}\equiv a/\Gamma$
is the expected radial velocity of a particle before ionization (and
so the radial dependence $r^{-2}$ is divided out here).

Since the lifetimes are exponentially distributed, they do not depend
on \textit{when} the estimate is made, the distribution of remaining
lifetime is always the same. In particular, since we have ionization
equilibrium, the lifetime probability distribution is equal to the
distribution of times \textit{since recombination}. The normalized
column density profile of particle density as a function of velocity
$N(v)$ is therefore equal to Eq.~\ref{eq:profile}: \begin{equation}
N(v)=v_{\mathrm{ion}}^{-1}\exp(-\frac{v}{v_{\mathrm{ion}}}).\label{eq:velprof0}\end{equation}

\subsection{Neutrals braked by ion gas}
\label{sub:brake}

If the neutrals moving radially outwards are interacting with a gas
at rest, the neutrals will be braked. The neutral-neutral interaction
cross-section is tiny, however, and requires a very high gas density
to be effective. Much more efficient is neutral-ion interaction, due
to the polarizability of neutral elements. Since the gas around $\beta$\,Pic
is strongly ionized, we now look closer on how the absorption profile
would be affected from the braking.

The braking of neutrals in an ion gas is analogous to ions being braked in a
neutral gas, a problem studied by \citet{beu89}. The equation
of motion for a neutral particle of mass $m$ accelerated by radiation
and braked by an ion gas is

\begin{equation}
m\frac{\mathrm{d}v}{\mathrm{d}t}=\beta\frac{GM_{\star}m}{r^{2}}-kv,\label{eq:motion}\end{equation}

where\begin{equation}
k=\pi\sqrt{\frac{4\alpha e^{2}}{4\pi\varepsilon_{0}}}\sum_{i}n_{i}\frac{m_{i}}{\sqrt{\mu_{i}}},\end{equation}

$n_{i}$ is the number density of ions of species $i$, $m_{i}$ is
the ion mass, $\mu_{i}=m_{i}m/(m_{i}+m)$ is the reduced mass, $\alpha$
is the polarizability of the neutral particle, $e$ is the electron
charge, and $\varepsilon_{0}$ is the electrical permettivity. The
asymptotic stable solution is\begin{equation}
v=v_{\mathrm{drift}}\equiv\frac{\beta GM_{\star}m}{kr^{2}}.\end{equation}
If $v_{\mathrm{drift}}t_{\mathrm{drag}}\ll r$, where $t_{\mathrm{drag}}\equiv m/k$,
then the time-dependent solution is \begin{equation}
v(t)\approx v_{\mathrm{drift}}\left[1-\exp(-t/t_{\mathrm{drag}})\right],\end{equation}
where we have assumed that the particle starts without initial velocity.
The probability distribution for the lifetime $\Lambda$ of a neutral
particle is still given by Eq.~\ref{eq:lifetime}, so the resulting
velocity distribution $v(\Lambda)$ is described by the probability
density distribution
\begin{equation}
N(v)=f_{\Lambda}\left(t(v)\right)\left|\frac{\mathrm{d}t(v)}
{\mathrm{d}v}\right|=v_{\mathrm{ion}}^{-1}\left(1-\frac{v}{\gamma v_{\mathrm{ion}}}\right)^{(\gamma -1)},\label{eq:velprof1}
\end{equation}
where $\gamma \equiv v_{\mathrm{drift}}/v_{\mathrm{ion}} = \Gamma t_{\mathrm{drag}}$.
For $\gamma\gg1$, Eq.~\ref{eq:velprof1} reduces
to Eq.~\ref{eq:velprof0}. The expected velocity is
\begin{equation}
E\left[v\right]=\frac{\gamma v_{\mathrm{ion}}}{\gamma+1},
\end{equation}
which turns to
\begin{equation}
E\left[v\right]\approx v_{\mathrm{ion}}
\end{equation}
for $\gamma\gg1$. From Eq.~\ref{eq:velprof1}
follows that there generally is a radial dependence of the velocity
profile (as it depends on local circumstances, such as ion gas density),
in contrast to the freely accelerating profile of Eq.~\ref{eq:velprof0}.

\subsection{Self-shielding: numerical model}

The photons that contribute to the acceleration of the gas are generally
scattered in a direction different from the line of sight. This extinction
implies that gas located further out will see fewer photons, resulting
in a lower effective radiation force coefficient $\beta$; this is
called \emph{self-shielding}. The coefficient $\beta$ for a particle
will thus depend on the column density of particles between its position
and the star, of particles that share the same velocity. The $\beta$
coefficient then in turn determines the acceleration, and what velocity
the particle will reach before getting ionized. Because $\beta$ is
changing with velocity, and depends on the radial density profile
of particles, we model this effect of self-shielding numerically.
The numerical model is simplified by the one-dimensional nature of
the problem, and that gas at outer radii are affected by gas inside,
but not the reverse. For modeling the absorption profile from $\beta$\,Pic,
we assume the density, ionization, and temperature distribution computed
by \citet*[hereafter \citetalias{zag10}]{zag10}. The procedure adopted as follows:
\begin{enumerate}
\item Divide the radial distance into $N_{\mathrm{shell}}$ shells of thickness
$\Delta r$. Start with the innermost shell and compute the velocity
distribution of particles in each shell going outwards.
\item \label{enu:iter}Use $N_{\mathrm{part}}$ number of particles in a
shell (where each particle represents a number of real particles).
Initialize all particles with a random thermal velocity $v$ according
to the probability distribution $V_{\mathrm{T}}(v)=(2\pi\sigma_{\mathrm{T}}^{2})^{-1/2}\exp\left[-0.5(v/\sigma_{\mathrm{T}})^{2}\right]$,
where $\sigma_{\mathrm{T}}=\left[k_{\mathrm{B}}T_{i}/m\right]^{1/2}$,
$T_{i}$ is the temperature of the shell $i$, and $m$ is the mass
of the particle. The radius of shell $i$ is $r_{i}$.
\item Use the velocity dependent optical depth $\tau(v)$ towards the star
to compute $\beta$ as a function of $v$. There is no extinction
inside the first shell.
\item Iterate $N_{\mathrm{iter}}$ times. For each iteration, evolve the
velocity of each particle according to the equation of motion of Eq.~\ref{eq:motion}
using a time step of $\Delta t\propto\Gamma_{\mathrm{ion}}^{-1}(r_{i})$,
with the addition that $\beta$ is a function of $v$. For each iteration
and particle, reset the particle velocity to the thermal velocity
with probability $\Gamma_{\mathrm{ion}}\Delta t$. This simulates
that the particle gets ionized. Since we assume ionization equilibrium,
there is a recombination for each ionization, and the recombined particle
is set to the thermal velocity of the gas (the assumption is that
ions are efficiently braked to on average $\sim$zero radial velocity).
After $N_{\mathrm{iter}}$ iterations the solution has converged to
a stable statistical distributions of radial velocities of the particles.
\item Make a histogram of the velocities of all particles inside the present
shell, with $N_{\mathrm{vel}}$ velocity bins and bin size $\Delta v$.
Normalize the distribution such that it corresponds to the total column
density of the shell, $\sum_{v_{j}}\Delta N_{i}(v_{j})=n_{\mathrm{gas}}(r_{i})\Delta r$. 
\item Compute the extinction towards the star for the next shell by adding
the velocity dependent column density for each shell, $N_{i}(v_{j})=\sum_{k\leq i}\Delta N_{k}(v_{j})$,
and find $\tau_{i}(v_{j})=N_{i}(v_{j})\sigma_{0}c/(\lambda_{0}\Delta v)$,
where $\lambda_{0}$ is the wavelength of the transition and \begin{equation}
\sigma_{0}=\frac{A_{ij}\lambda_{0}^{4}}{8\pi c}\frac{g_{j}}{g_{i}}\end{equation}
 is the cross-section times wavelength per particle for the transition,
where $A_{ij}$ is the Einstein coefficient and $g_{j}$ and $g_{i}$
the statistical weights of the levels.
\item Repeat from (\ref{enu:iter}) for all shells. The observed profile
can then be estimated from $\tau_{N_{\mathrm{shell}}}(v)$.
\end{enumerate}
We implemented this model in \textsc{matlab} for \ion{Na}{1} around $\beta$\,Pic,
and verified it by testing against conditions where the analytical
profiles of \S\S\,\ref{sub:free} and \ref{sub:brake} should be valid,
with excellent agreement. In detail, we used $N_{\mathrm{shell}}=186$
with $r_{i}$ between 15 and 200\,AU (and thus $\Delta r=1$\,AU),
$N_{\mathrm{part}}=10^{4}$ per shell,
$N_{\mathrm{iter}}=10^{4}$ iterations, and a time step of $\Delta t=[2000\Gamma_{\mathrm{ion}}(r)]^{-1}$
(meaning we iterate 5 times longer than the ionization time scale).
For the velocity grid, we used $\Delta v=100$\,m\,s$^{-1}$ from
$-2$ to 20\,km\,s$^{-1}$, to be certain to cover most of the particle
velocities. For the \ion{Na}{1} D$_{2}$ line $\sigma_{0}=1.987\times10^{-13}$\,cm$^{2}$\,Å\,atom$^{-1}$,
and for the D$_{1}$ line the cross section is 1/2 as much. Since
these two resonance lines are completely dominating the radiation
force on \ion{Na}{1}, we computed the optical depth $\tau(v_{j})$ for the D$_{2}$
line and the effective $\beta(v_{j})=\beta_{\mathrm{Na\, I}}(\exp[-\tau(v_{j})]2/3+\exp[-\tau(v_{j})/2]/3)$,
where $\beta_{\mathrm{Na\, I}}$ is the radiation force coefficient
on \ion{Na}{1} without extinction.

\section{HARPS data}
\label{sec:Data}

To test how well the ion braking scenario compares to observations,
we retrieved 489 spectra of $\beta$\,Pic from the ESO archive. The
spectra were obtained by HARPS at the ESO 3.6\,m telescope at La Silla,
Chile, during the period of 2003-10-28 to 2008-03-21. HARPS is a fiber-fed
high-resolution ($R\sim10^{5}$) echelle spectrograph ($\lambda=3800-6900$\,\AA),
dedicated to the search for exoplanets. The long-term radial velocity
accuracy of HARPS is estimated to be $\lesssim1$\,m\,s$^{-1}$
\citep{may03}. The HARPS archive at ESO contains data processed by the
HARPS science-grade reduction pipeline, producing high-quality extracted
wavelength-calibrated spectra. Our main interest are regions around
the \ion{Ca}{2} H \& K doublet at 3934\,\AA\ and 3968\,\AA, the \ion{Fe}{1} 3860\,\AA\ 
line, and the region around the \ion{Na}{1} D doublet at 5890\,\AA\ and 5896\,\AA\ 
(see Table\,\ref{tab:Lines} for line details). Typical signal to
noise ratios in single spectra range from 50--100 for the 3900\,\AA\ region, 
to 100--250 around 5900\,\AA, depending on exposure time and seeing.
\begin{deluxetable}{cclccc}
\tablecaption{Lines studied\label{tab:Lines}}
\tablehead{\colhead{Element} & \colhead{Design.} & \colhead{$\lambda_{\mathrm{air}}${[}\AA{]}} &
\colhead{$A_{ul}${[}s$^{-1}]$} & \colhead{$g_{u}-g_{l}$} & \colhead{Ref.\tablenotemark{c}}}
\startdata
\ion{Ca}{2} & K & 3933.6614 & $1.47\times10^{8}$ & $4-2$ & 1\\
\ion{Ca}{2} & H & 3968.4673 & $1.4\times10^{8}$ & $2-2$ & 1\\
\ion{Fe}{1} &  & 3824.4436 & $2.83\times10^{6}$ & $7-9$ & 1\\
\ion{Fe}{1} &  & 3859.9114 & $9.70\times10^{6}$ & $9-9$ & 1\\
\ion{Na}{1} & D$_{2}$ & 5889.950954 & $6.16\times10^{7}$ & $4-2$ & 2\\
\ion{Na}{1} & D$_{1}$ & 5895.924237 & $6.14\times10^{7}$ & $2-2$ & 2\\
\enddata
\tablenotetext{c}{Wavelength references are (1) \citet{mor03}, (2) \citet{wol08}}
\end{deluxetable}

\subsection{Cleaning telluric lines}
\label{s:clean}
With the high-quality reduced data produced by the HARPS pipeline,
not much post processing needs to be done, in general. The exception
is the \ion{Na}{1} absorption region at 5890\,\AA, where abundant time-varying
absorption features from atmospheric H$_{2}$O interfere with the CS
absorption lines of interest. Common practice is to remove the telluric
features by observing a nearby {}``telluric standard'' right before
or after the science observation, and divide out the features
\citep[e.g., ][]{vac03}. Another is to fit a theoretical atmosphere
spectrum \citep{sei10}. Since we did not have
access to a standard star and it is hard to make a good model fit without
leaving systematic residuals, we opted for a third method, making
use of the strengths of this particular data set: 
\begin{enumerate}
\item The supreme stability of HARPS (including the line-spread function).
\item The large number of observations (489).
\item The fact that all telluric absorption lines in the region are optically
thin.
\item The variation in position and strength of the telluric features, due
to the variable barycentric correction from the orbital and rotation
speed of the Earth and the variable airmass and water content of the
atmosphere.
\end{enumerate}
By assuming that the stellar spectrum is static and that only the
strength and relative shift of the telluric lines are varying, we
can use the large number of spectra to constrain the problem. To constrain
the telluric absorption lines even further, we downloaded multi-epoch
spectra for three additional A0 stars: HD\,71155 (40 spectra), HD\,177724
(12), and HD\,188228 (24). The advantage of A0 stars is that they
have very weak \ion{Na}{1} features that are rotationally broadened to
several 100\,km\,s$^{-1}$, making them suitable as spectrally smooth
background lamps to study telluric absorption lines. The problem now
becomes to find a (normalized) telluric extinction spectrum $\kappa(\lambda)$,
a model spectrum $F_{\mathrm{mod}}^{k}$($\lambda)$ (for each star
$k$), and a water column parameter $\alpha_{k,n}$ (for each observation
$k,n$) such that 
\begin{eqnarray}
\chi_{\mathrm{mod}}^{2} = \sum_{k}\sum_{n=1}^{N_{k}}\int_{\lambda}
\frac{1}{\sigma_{\mathrm{obs}}^{k,n}(\lambda)}
|F_{\mathrm{obs}}^{k,n}(\lambda) - \nonumber \\
F_{\mathrm{mod}}^{k}([1+z_{k,n}]\lambda)\exp(-\alpha_{k,n}\kappa[\lambda])|
\mathrm{d}\lambda\label{eq:chi2}
\end{eqnarray}
 is minimized, where $N_{k}$ is the number of observations of star
$k$, $\sigma_{\mathrm{obs}}^{k,n}(\lambda)$ is the error of the
measurement of spectrum $k,n$ at wavelength $\lambda$, and $z_{k,n}$
is the (known) redshift associated with observation $k,n$. We implemented
a non-linear conjugate gradient minimization method in C++, using the
Polak-Ribi\`{e}re method with restart to guarantee convergence
\citep[see, e.g.,][]{she94}. Since
$\chi^{2}$ minimization in general is degenerate with respect to
a continuous spectrum (i.e.\ many different spectra can produce the
same minimum $\chi^{2}$), we added a second-order ``smoothness'' regularization
that penalizes rapid changes of the slope:
\begin{equation}
\chi_{\mathrm{reg}}^{2}(\{f_{m}\})=\sum_{m=1}^{M}\left(\frac{f_{m-1}+f_{m+1}}{2}-f_{m}\right)^{2},
\label{eq:reg}
\end{equation}
 where $f_{m}$ is the value of pixel $m$ of the discrete model spectrum
(or model extinction) of $M$ pixels, and $f_{0}=f_{M-1}$ are boundary
values (arbitrarily fixed to 1 for stellar spectra and 0 for extinction).
In practice we thus minimize 
\begin{equation}
\chi^{2}=\chi_{\mathrm{mod}}^{2}+a\chi_{\mathrm{reg}}^{2}
(\{\kappa_{m}\})+b\sum_{k}\chi_{\mathrm{reg}}^{2}(\{F_{\mathrm{\mathrm{mod},}m}^{k}\}),
\end{equation}
 where $a$ and $b$ are regularization constants that decide the
relative importance of the regularization. These constants have to
be manually tuned to produce good results; too small and the minimization
process will produce numerically noisy spectra, too high and the spectra
will be too smooth, devoid of detail. Fortunately, there is a large
range between the too small and the too high, that produce good spectra
that only weakly depend on the regularization constants.
We found that $a = b = (10-1000)\times N_{\mathrm{spec}}$, where $N_{\mathrm{spec}}$ 
is the total number of spectra used, produced good results; 
in the end, we used $a = b = 250\times N_{\mathrm{spec}}$.

\begin{figure}
\plotone{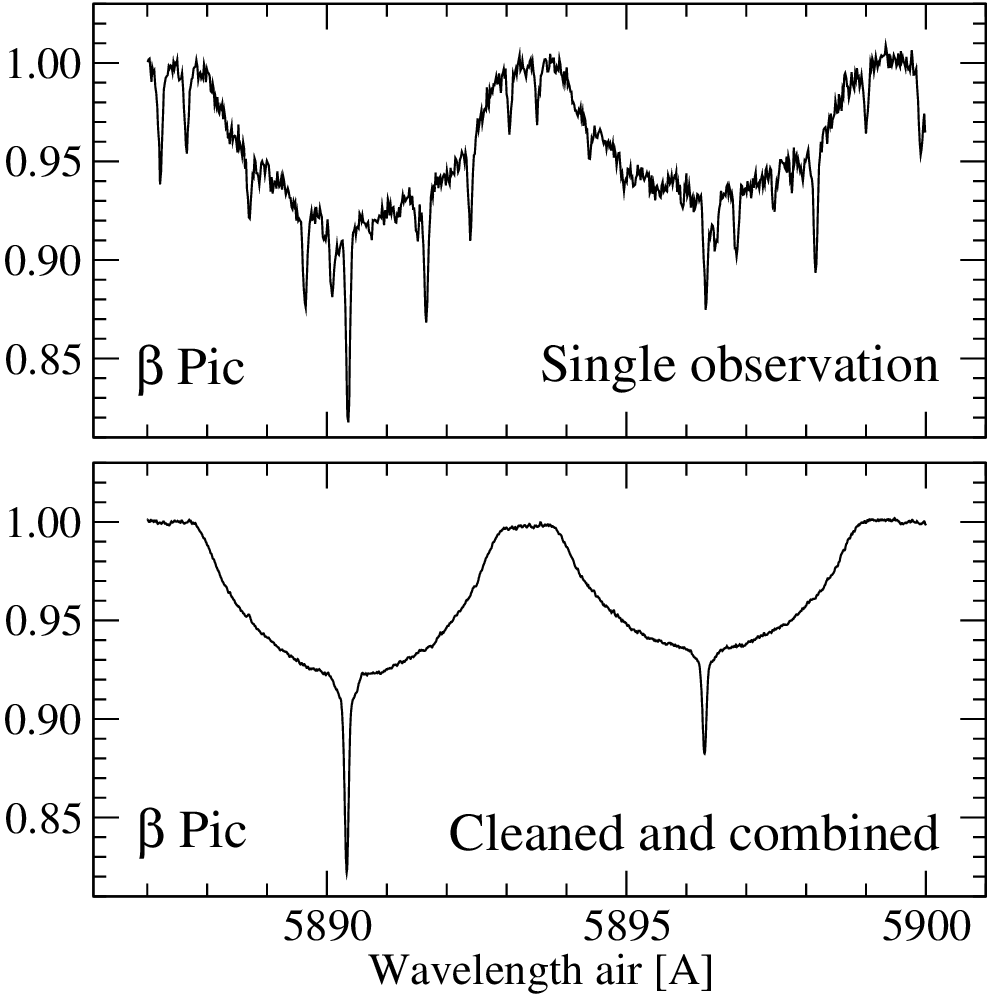}
\caption{The upper panel shows a single spectrum of $\beta$\,Pic
as obtained from the HARPS pipeline around the \ion{Na}{1} D lines.
This region is plagued by numerous telluric absorption lines from H$_2$O.
The lower panel shows the resulting spectrum from a combination of 92
FEB-free spectra, where the tellurics have been cleaned (as described in
\S\,\ref{s:clean}).\label{f:cleanNa}}
\end{figure}

After an optimal fit to the telluric lines were found (simultaneously
with the average stellar spectra), we divided (``cleaned'') all individual
observations by the fitted telluric extinction. This allows us to inspect 
individual clean spectra and accommodate potential changes between
epochs. The improvement from the cleaning process is illustrated in
Fig.~\ref{f:cleanNa}.

\subsection{Deriving line profiles}
\label{sec:linederiv}
\begin{figure}
\plotone{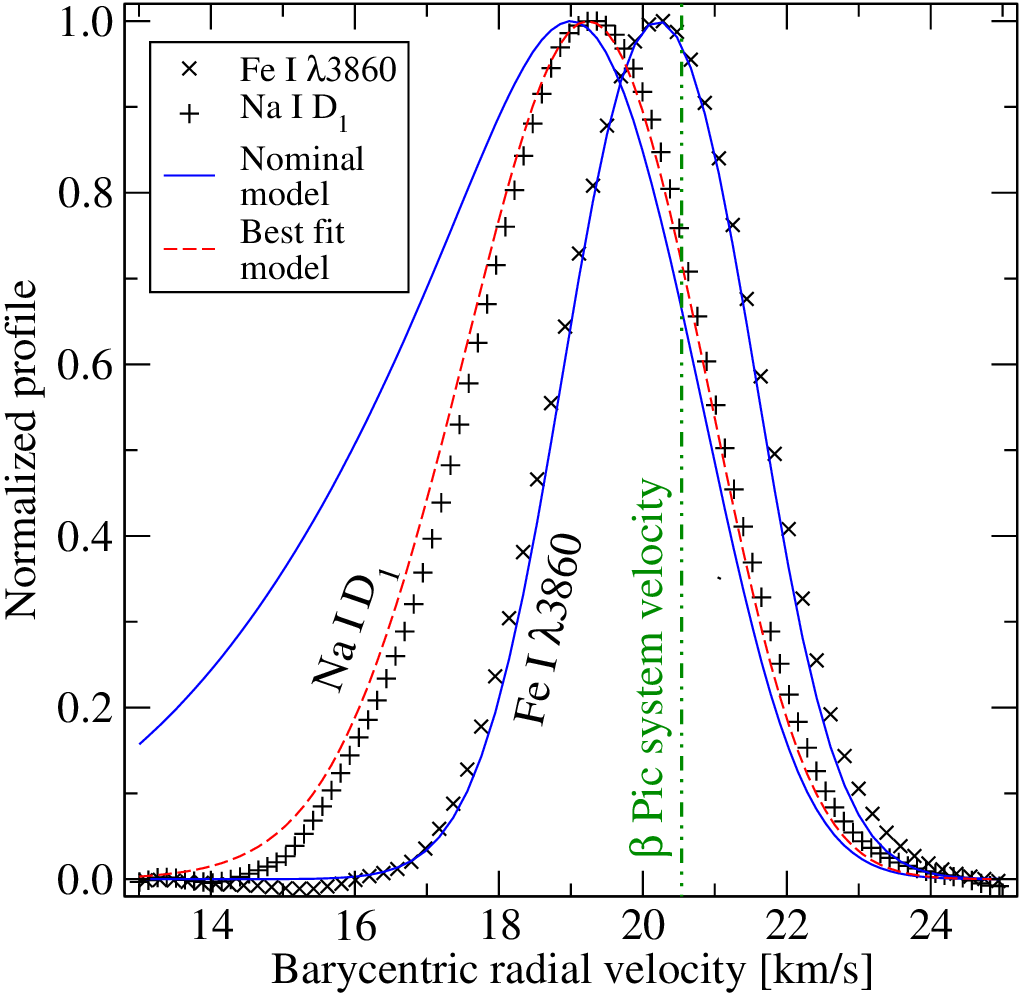}
\caption{The observed normalized line profiles of \ion{Na}{1} D$_1$ and
\ion{Fe}{1} $\lambda$3860 with over plotted models. The crosses show the observed
\ion{Fe}{1} line profile, the plus signs the observed \ion{Na}{1} line, and
the solid lines are from the nominal model (described in \S\,\ref{sec:methods}). 
As seen, the blue wing of the 
\ion{Na}{1} line is greatly overestimated by the model compared to the observed
profile, while the \ion{Fe}{1} line looks fine. The dashed line shows the best-fit
model for \ion{Na}{1}, where the carbon abundance has been increased by a factor
of 2 compared to the nominal case (the \ion{Fe}{1} profile does not change; see
\S\ref{sec:results}). The expected errors on the measured profiles are 1\,\% of the peak for \ion{Na}{1} and 3\,\% of the peak for \ion{Fe}{1}
(see \S\,\ref{sec:linederiv}).
\label{f:lineprof}}
\end{figure}
In addition to the stable gas absorption lines seen against $\beta$\,Pic,
some lines are known to show strongly velocity-shifted transient absorption
features \citep{lag87}. These absorption features are thought to be originating
from ``falling evaporating bodies'' (FEBs) close to the star \citep{beu89},
and are mostly observed in the UV \citep{del93} and in the \ion{Ca}{2} H and
K lines. Their relevance to the present work is that we want to avoid
spectra with strong FEB features, as they may bias the line profiles
we are studying. Fortunately, neither \ion{Na}{1} nor \ion{Fe}{1} are known to
show FEB features, but that is likely a sensitivity issue. Since Ca,
Na, and Fe around $\beta$\,Pic are strongly ionized \citepalias{zag10},
the column density of \ion{Ca}{2} is much higher than either \ion{Na}{1} or
\ion{Fe}{1}, making the \ion{Ca}{2} H and K lines strongly saturated and FEB
features in those lines stronger, despite having oscillator strengths
similar to the \ion{Na}{1} and \ion{Fe}{1} resonant transitions. To reduce the
possibility of FEBs affecting the \ion{Na}{1} and \ion{Fe}{1} line profiles,
we inspected all 489 spectra for FEB features in the sensitive \ion{Ca}{2}
lines, and selected 95 without any sign of FEB activity%
\footnote{In order to provoke a FEB signal in the \ion{Na}{1} lines, we also selected
a sample of 262 spectra that show strong red-shifted FEB activity
in \ion{Ca}{2}. Adding the frames, the resulting \ion{Na}{1} indeed shows a red-shifted FEB feature
with an equivalent width $0.8_{-0.4}^{+0.8}$\,m\AA, i.e.\ at the
$\sim$10\% level of the main stable line.%
}. With no FEBs visible in the \ion{Ca}{2} lines, we expect their contribution
to the \ion{Na}{1} and \ion{Fe}{1} lines to be negligible. Subregions of the
95 spectra around the \ion{Na}{1} 5890\,\AA, \ion{Fe}{1} 3824\,\AA, and 
\ion{Fe}{1} 3860\,\AA\ lines were
then extracted and interpolated to a velocity scale of 0.1\,km\,s$^{-1}$
resolution (using the HARPS wavelength calibration and line center
wavelengths from Table~\ref{tab:Lines}), stacked (weighted by their
signal to noise), and finally normalized. The resulting profiles for
\ion{Fe}{1} 3860\,\AA\ and \ion{Na}{1} 5890\,\AA\ are shown in Fig.~\ref{f:lineprof}. As seen
in the figure, there is a significant offset between the observed
radial velocities of \ion{Na}{1} and \ion{Fe}{1}, as expected from the ion-braking
scenario.

Given the large number of high signal-to-noise spectra being combined, errors
in the derived profiles are expected to be dominated by systematic errors, such
as flat-fielding, flux calibration, and telluric line modeling errors. To check 
the reliability of the derived profiles, we compared two lines from the same
element with each other; many of the expected systematic effects should change with
wavelength. One notable exception is contamination in the line profile due to FEB
activity; since only spectra without FEBs visible in the \ion{Ca}{2} are selected,
we estimate the impact on the  \ion{Fe}{1}/\ion{Na}{1} regions to be $<$\,1\%.
By comparing the \ion{Fe}{1} 3824\,\AA\ with the 3860\,\AA\ line, we find them
to be consistent to within 3\,\% (1 standard deviation) of the peak. Making the 
same comparison between the \ion{Na}{1} 5890\,\AA\ and  5896\,\AA\ lines, we find
them to be within 1\,\% of each other. The relative radial velocity error is negligible
in comparison to the line profile; it is dominated by the 10\,m\,s$^{-1}$ accuracy
by which the \ion{Fe}{1} lines are known (HARPS instrumental errors are
$\lesssim 1$\,m\,s$^{-1}$). This means that the velocity difference between the 
\ion{Fe}{1} and \ion{Na}{1} lines, which from Fig.~\ref{f:lineprof} is seen to be on the order 
of 1\,km\,s$^{-1}$, is 100$\times$ larger than the estimated velocity uncertainty, 
and therefore highly significant.

\subsection{Removing interstellar absorption}
\label{s:IS}
\begin{figure}
\plotone{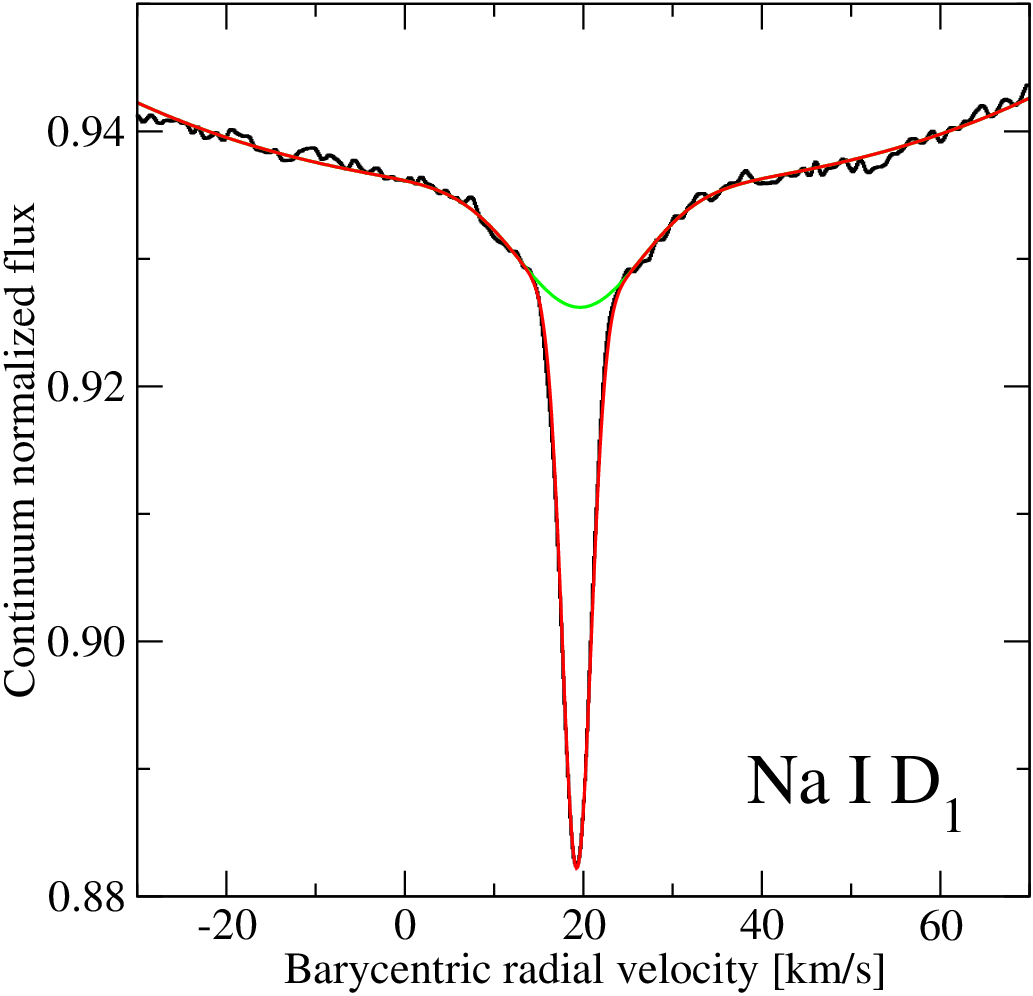}
\caption{The central part of the \ion{Na}{1} D$_1$ line. The curves over plotted
the data are the rotational+IS+CS profile, and the rotational+IS profile (used
to fit the continuum for the CS line). See \S\,\ref{s:IS}. 
\label{f:linefit}}
\end{figure}
The narrow CS absorption lines in Fig.~\ref{f:cleanNa} 
are seen to have ``shoulders'', which are stable in time and due
to interstellar (IS) extinction \citep{vid86}. To remove them from the 
profile, we assumed they were Gaussian shaped, and simultaneously
fit a rotational profile with two Gaussians per Na line, one for 
the the CS and IS contribution, respectively (Fig.~\ref{f:linefit}). 
The rotational profile was best fit
by $v\sin i = 126 \pm 1$\,km\,s$^{-1}$ and a limb-darkening coefficient 
$\epsilon = 0.57 \pm 0.06$. The ratio
between the two IS D$_2$ and D$_1$ line strengths is 2.0, indicating
optically thin absorption. The IS lines are centered at a heliocentric
radial velocity of $19.46 \pm 0.27$\,km\,s$^{-1}$ (i.e.\ close to the
stellar velocity at $20.5$\,km\,s$^{-1}$) and are unusually broad
with a broadening parameter $b = 21.4 \pm 1.7$\,km\,s$^{-1}$, related to 
the standard deviation $\sigma$ of the normal 
distribution through $b = 2^{1/2}\sigma$. \citet{vid86} found the IS line
to be present also in the nearby star $\alpha$\,Pic and were able to remove it
from their $\beta$\,Pic spectrum (along with telluric lines) by dividing the
$\beta$\,Pic spectrum with that from $\alpha$\,Pic. Since we do not have access
to an $\alpha$\,Pic spectrum of a quality similar to the $\beta$\,Pic spectra,
we remove the IS line by division with the fitted Gaussian.

\subsection{Estimating the line-spread function}

To compare the theoretical line profile models with the data, we need
to convolve the models with the instrumental profile of the spectrograph,
the line-spread function (LSF). One way to estimate the LSF is to
measure the shapes of spectral lines from a Th-Ar lamp, since they
are expected to be unresolved. Because HARPS records Th-Ar spectra in
a parallel fiber with (almost) all science observations, we have a
huge data set of spectral lines to choose from. Since we are interested
in spectral regions around 3900\,\AA\ and 5900\,\AA, we decided to determine
the LSF for those two regions separately by only selecting suitable
lines (unblended lines of sufficient flux that do not come close to
saturate) from the respective orders. With over 500 spectra retrieved
from the archive, we ended up with several thousand lines for each
region. The lines were then linearly interpolated and shifted to a
grid of 0.1\,km\,s$^{-1}$ resolution, using the wavelength calibration
provided by the HARPS pipeline. Finally, the lines were added, weighted
by the signal to noise, fit by a third-degree polynomial (for the
background) + a double Gaussian of the functional form
\begin{equation}
\mathrm{LSF}(v)=\sum_{i=1,2} k_{i}\exp\left(-\frac{\left[v-m_{i}\right]^{2}}{2\sigma_{i}^{2}}\right)
\label{eq:lsf}
\end{equation}
 where $v$ is the velocity shift from the center of the line, and
the other constants are fitted. Comparing the LSFs determined for
the 3900\,\AA\ and 5900\,\AA\ regions, we found no significant difference
($<$1\% of the peak). The fitted parameters $k_{1}=1.193$, $m_{1}=5.64$\,m\,s$^{-1}$,
$\sigma_{1}=1.135$\,km\,s$^{-1}$, $k_{2}=-0.194$, $m_{2}=-9.01$\,m\,s$^{-1}$,
and $\sigma_{2}=0.680$\,km\,s$^{-1}$ approximate the LSF very
well, with residuals again below 1\% of the peak. The full-width half
maximum (FWHM) of the LSF is 2.67\,\,km\,s$^{-1}$, meaning a spectral
resolution of $R\sim112\,000$.

\section{Analytic methods}
\label{sec:methods}
\begin{figure}
\plotone{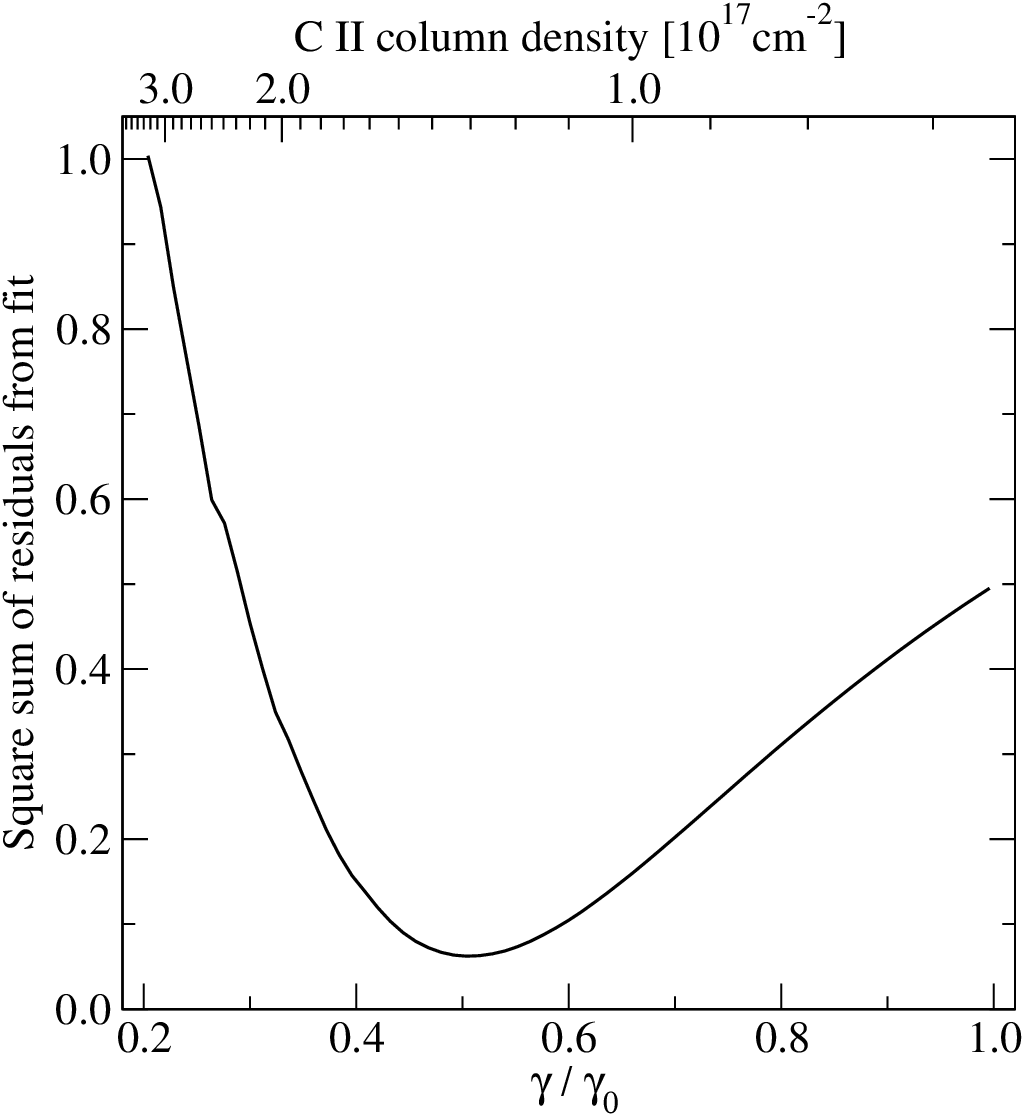}
\caption{Using a model where the ionization velocity $v_{\mathrm{ion}}$
is reduced as much as possible while still being barely consistent with 
observations, we here plot the square sum of the residuals between a
fit of a particular $\gamma$ as a function of $\gamma/\gamma_0$, with $\gamma_0$
being the nominal value. The corresponding column densities are given on
the upper horizontal axis.\label{f:chi2}}
\end{figure}
With a stellar mass of $M_{\star}=1.75\,M_{\odot}$ \citep{cri97},
\citetalias{fer06} compute $\beta_{\mathrm{Na\, I}}=360\pm20$, $\beta_{\mathrm{Fe\, I}}=27\pm2$,
$\Gamma_{\mathrm{Na\, I}}(100\,\mathrm{AU})=1.1\times10^{-7}$\,s$^{-1}$ and
$\Gamma_{\mathrm{Fe\, I}}(100\,\mathrm{AU})=5.8\times10^{-8}$\,s$^{-1}$, resulting
in $v_{\mathrm{ion}}^{\mathrm{Na\, I}}=3.3$\,km\,s$^{-1}$ and
$v_{\mathrm{ion}}^{\mathrm{Fe\, I}}=0.5$\,km\,s$^{-1}$. As our nominal gas
disk model, we use the $\beta$\,Pic model with parameters described in 
\citetalias{zag10}. Assuming
free acceleration, the associated line profiles of Eq.\,\ref{eq:velprof0}
are adjusted for the radial velocity of $\beta$\,Pic by fitting
the \ion{Fe}{1} profile (as seen in Fig.~\ref{f:lineprof}; the fitted
radial velocity of the system is $20.5\pm0.2$\,km\,s$^{-1}$). The theoretical 
profile looks
fine for \ion{Fe}{1} and the red wing of \ion{Na}{1}, but the modeled 
\ion{Na}{1} blue wing is clearly more extended than the observed. Apparently
the Na atoms of the $\beta$\,Pic disk do not reach as high speeds as we 
expect before getting ionized. The discrepancy between model and observations 
could be due to the following three reasons:
\begin{enumerate}
\item Less acceleration, e.g., an overestimated $\beta$ value for \ion{Na}{1}.
Since the $\beta$\,Pic spectrum is very well constrained around the 
\ion{Na}{1} D lines, however, and the corresponding radiative 
transition coefficients are known to very high accuracy for these resonance
lines, it is unlikely that the $\beta$ value is more off than the estimated 
8\%.
\item Higher ionization rate of \ion{Na}{1}. This will give less time for
the atom to accelerate before getting ionized. The ionization rate is
directly proportional to the ionizing UV radiation from the star. The UV
part of the $\beta$\,Pic spectrum is not well constrained outside the
spectral windows observed by FUSE (905--1187\,\AA) and HST/STIS (1459--2888\,\AA), 
but most of the ionizing
photons are expected to arise from close to the ionization threshold
at 2413\,\AA, where the spectrum is known to within 10\% from STIS
\citetext{Aki Roberge, priv.\ comm.\ 2010}. The ionization
model of \citetalias{zag10} reproduce observed ionization ratios
in the $\beta$\,Pic gas disk reasonably well, so a much different
ionization rate than the assumed would be surprising.
\item More efficient braking. With stronger braking, the
resulting drag velocity is smaller. The braking is the strongest
for the higher speed particles, possibly explaining the observed absence
of a strong blue wing. Since the braking is dominated by, 
and directly proportional to, the number density of \ion{C}{2},
an underestimate of \ion{C}{2} would result in a proportional 
underestimate of the braking force. Strongly saturated
absorption lines of \ion{C}{2} observed by FUSE \citepalias{rob06}
have determined the C abundance in the $\beta$\,Pic disk to be at
least 20$\times$ overabundant with respect to other elements in the
disk at cosmic abundance. Since the stable CS lines are saturated and 
blended with a time-variable broad component due to FEBs, small 
variations in the intrinsic line shape can mimic large abundance
changes. It thus seems motivated to explore the possibility that 
the carbon is even more overabundant than previously thought, and
what consequences that has for the braking and the line profile of 
\ion{Na}{1}.
\end{enumerate}
A combination of the above is also possible. In particular, should 
braking be more efficient than anticipated, the lower radial velocities
will contribute to stronger self-shielding and a decrease in the 
$\beta$-value, resulting in less acceleration.

To investigate the possible scenarios producing the observed line
profiles, we made an extensive parameter search for $v_{\mathrm{ion}}$
and $\gamma$, trying to fit the profile from Eq.~\ref{eq:velprof1} to 
the observations 
(leaving the system radial velocity as a free parameter). Because
self-shielding, as explored in the numerical model, in most
of the parameter space is very small (with an optical
depth less than 0.1 resulting in only a minor correction to
the analytical profile), we only used the numerical model
to investigate a subset of models where self-shielding was expected
to be most important.

\section{Results and discussion}
\label{sec:results}
Since the expected radial velocity increases with both $v_{\mathrm{ion}}$
and $\gamma$, and since the observed line profiles are not well resolved, the
problem of finding the optimal $v_{\mathrm{ion}}$ and $\gamma$ becomes
ill-conditioned. Even so, we can constrain the parameter space by 
finding what combinations of parameters are consistent with the line
profiles. One question is if additional braking is necessary, or if a
moderation of $v_{\mathrm{ion}}$ by stretching known quantities within
their estimated uncertainties is sufficient to explain the profile. This
should put a lower limit on the extra braking needed, and by extension
put an independent lower limit on the \ion{C}{2} content in the disk. We
get a lower reasonable limit for the $v_{\mathrm{ion}}$ of \ion{Na}{1} by assuming 
$\beta = 330$ (which is 1.5\,$\sigma$ below its estimated value) and an
ionization rate that is 20\,\% higher than the nominal, resulting in
$v_{\mathrm{ion}} = 2.5$\,km\,s$^{-1}$. In Fig.~\ref{f:chi2} the least square-sum
of the residuals between the observed line profiles for \ion{Na}{1} 
and \ion{Fe}{1}, and their models (where the system radial velocity again 
was a free parameter) are plotted as a function of $\gamma/\gamma_0$, where
$\gamma_0$ corresponds to the $\beta$ and $\Gamma$ assumed above, and a braking 
ion gas density according to the model by \citetalias{zag10}. The corresponding column density
of \ion{C}{2} is given on the upper horizontal axis of the figure. The
smallest residuals are found for  $\gamma/\gamma_0 = 0.5$, corresponding to
the \ion{C}{2} column density $N_{\mathrm{C\,II}} = 1.3\times10^{17}$\,cm$^{-2}$,
which is 2$\times$ the model column density of \citetalias{zag10}, and 6$\times$
the $N_{\mathrm{C\,II}} = 2.0^{+2.1}_{-0.4}\times10^{16}$\,cm$^{-2}$ reported
by \citetalias{rob06}. The best-fit \ion{Na}{1} velocity profile resulting
from increased braking is shown in Fig.~\ref{f:lineprof}. Because the
braking is proportional to the velocity, only the high-velocity tail of
the profile is significantly affected by the \ion{C}{2} density increase.
Similarly, \ion{Fe}{1} never gets accelerated to high enough velocities
for braking to be important, so the \ion{Fe}{1} velocity profile does
not change with the increased \ion{C}{2} density.

The implicated overabundance of C is on the order of 100$\times$
compared to cosmic abundances of other detected elements. Perhaps this
reflects the intrinsic abundance of the gas production mechanism; e.g., 
\citet{kar03} find carbon rich FEBs essential to explain the high-velocity
FEB spectral features. Another possibility is that the current
abundances are more strongly dependent
on the gas \textit{removal} mechanism. Even with an efficient ion braking
mechanism, any specific particle is neutral some fraction of the time. If the
radiation pressure on the particles in their neutral state overcomes gravity, 
they will be accelerated outwards until they get ionized (and braked again).
This will effectively remove elements at a rate proportional to their
ionization speed $v_{\mathrm{ion}}$ and neutral fraction. On 
the other hand, elements that do not experience sufficient radiation pressure in their 
neutral phase \citepalias[i.e., H, He, C, N, O, F, Ne, and Ar, see Table 1 of][]{fer06}
are thus expected to pile up, thereby increasing their relative abundance compared
to their production abundance. Unfortunately, the elements that experience little
radiation pressure are also the elements that are difficult to observe, so only
C and O of that group have so far been detected \citepalias{rob06}, although H has observed 
upper limits \citep{fre95,lec01}.
Surprisingly, absorption line observations by FUSE favor a cosmic
abundance of disk \ion{O}{1} gas \citepalias{rob06}. Again, this conclusion is
sensitively dependent on the assumed (unresolved) line profile, as can be seen
in Fig.~1 of the \citetalias{rob06} online supplementary material: reducing the broadening 
parameter from the best-fit value of $\sim$14\,km\,s$^{-1}$ to a few km\,s$^{-1}$
results in a much higher, but less constrained, column density. A broadening parameter
as large as 14\,km\,s$^{-1}$ for the stable disk gas is unexpected in itself considering
other observed absorption lines, and is likely caused by multiple unresolved velocity
components from FEBs (as seen for other elements).

One way to confirm the overabundance of \ion{C}{2} around $\beta$\,Pic is to
observe its thermal 157.7\,$\mu$m emission, as the emission is predicted to be roughly
proportional to the C gas mass \citepalias{zag10}. The \textit{Infrared Space Observatory}
(ISO) observed $\beta$\,Pic with the LWS instrument and found a 4\,$\sigma$ emission
feature with an integrated flux of 10$^{-13}$\,erg\,s$^{-1}$\,cm$^{-2}$ \citep{kam03}.
This is $15\times$ more emission than predicted from a gas disk with 20$\times$ overabundant C
\citepalias{zag10}, and would imply a C abundance of $>$100$\times$ over cosmic values (compared
to other observed elements in the disk). Recently, the infrared space telescope 
\textit{Herschel} with the
PACS instrument observed both \ion{C}{2} 157.7\,$\mu$m and \ion{O}{1} 63.2\,$\mu$m
emission, confirming the ISO detection and finding that \ion{O}{1} probably also is
overabundant. The data are still being analyzed, however, as e.g.\ the
spectrophotometric calibration is not settled yet. A publication presenting the 
Herschel observations is in preparation by the key project
\textit{Stellar Disk Evolution} team lead by G.\,Olofsson.

The processes that produce the C-rich gas disk around $\beta$\,Pic must
reasonably be present also in other debris disk systems. The frequency and
efficiency of the process is not known, but the absence of detectable amounts
of gas around AU\,Mic \citep{rob05}, another edge-on debris disk system in the
$\beta$\,Pic moving group, shows that the process is not ubiquitous. A model to 
quantitatively investigate the balance of gas production and removal mechanisms 
in debris disks will be the subject for future studies.

\section{Concluding summary}
\label{sec:Conclusions}
The prediction made by the braking model of \citetalias{fer06},
that \ion{Fe}{1} and \ion{Na}{1} should show different radial
velocities, is confirmed. This shows that, to a first order
approximation, neutral elements are not braked by the gas,
while the ions must be. Looking more closely at the line
profiles, we see an absence of a \ion{Na}{1} high-velocity population, 
expected from free acceleration. Our
interpretation is that the high-velocity Na atoms must be
braked. A factor of 2--5$\times$ more \ion{C}{2} gas than
previously estimated is required for \ion{C}{2} to be the
braking agent; the implication is that the gas disk around 
$\beta$\,Pic is even more C rich than previously thought,
up to 100$\times$ cosmic abundance relative to other
detected elements.

\acknowledgments

The author acknowledges the useful discussions with Jean-Michel D\'{e}sert, 
Ricky Nilsson, G\"{o}ran Olofsson, Seth Redfield, Aki Roberge,
Philippe Th\'{e}bault, and Yanqin Wu. This work was supported by the 
\textit{Swedish National Space Board} (contract 84/08:1).

\bibliography{ms}

\end{document}